\documentclass[a4paper,10pt]{article}

\usepackage{bm}
\usepackage{latexsym}
\usepackage{dcolumn}
\usepackage{amsmath,amsfonts,amssymb}
\usepackage{graphicx,epsfig}
\usepackage{color}
\usepackage{verbatim}

\def\pl{Planck length }

\reversemarginpar

\title{Ideal Gas in a strong Gravitational field: Area dependence of Entropy}

\author{Sanved Kolekar\footnote{sanved@iucaa.ernet.in} ~ and T.~Padmanabhan\footnote{paddy@iucaa.ernet.in}\\
IUCAA, Pune University Campus, Ganeshkhind,\\
 Pune 411007, INDIA. \\ 
}

\date{}

\begin{document}


\maketitle

\begin{abstract}

 We study the thermodynamic parameters like  entropy, energy etc. of
 a box of gas made up of indistinguishable particles when the box is kept in various static background spacetimes having a horizon. 
We compute the thermodynamic variables using both statistical mechanics as well as by solving the hydrodynamical equations for the system.
When the box is far away 
  from the horizon, the entropy of the gas
depends on the volume of the box except for small corrections due to background geometry. As the box is moved closer to the horizon with one (leading) edge of the box at about Planck length ($L_p$) away from the horizon, the entropy shows an area dependence rather than a volume dependence. More precisely, it depends on a small volume $A_{\perp} L_p/2$ of the box, upto an order $ {\cal O} (L_p/K)^2$ where 
 $A_{\perp}$ is the transverse area of the box and 
$K$ is the (proper) longitudinal size of the  box related to the distance between leading and trailing edge in the vertical direction (i.e in the direction of the gravitational field).  Thus the contribution to the entropy comes from only a fraction $ {\cal O} (L_p/K)$ of the  matter degrees of freedom and the rest are suppressed when the box approaches the horizon. Near the horizon all the thermodynamical quantities behave as though the box of gas has a volume $A_{\perp} L_p/2$ and is kept in a Minkowski spacetime. These effects are: (i) purely kinematic in their origin and are independent of the spacetime curvature (in the sense that Rindler approximation of the metric near the horizon can reproduce the results) and 
 (ii) observer dependent. When the equilibrium temperature of the gas is taken to be equal to the the horizon temperature, we get the familiar $A_\perp / L_p^2$ dependence in the expression for entropy. All these results hold in a $D+1$ dimensional spherically symmetric spacetime. 
The analysis based on methods of statistical mechanics and the one based on thermodynamics applied to the gas treated as a fluid in static geometry, lead to the same results showing the consistency. The implications are discussed.
\end{abstract}

\section{Introduction} \label{sec:intro}

The first indication of the connection between thermodynamics and gravity came with the work of Bekenstein
 \cite{bekentropy, bekentropy2, bekentropy3} who proposed the idea that a black hole should have an entropy that is proportional to the area of its horizon. 
This interpretation became well established with the discovery of the temperature of the black hole \cite{hawking, hawking2}.
Soon afterwards, it was discovered that all horizons can be attributed a temperature \cite{Davies, gibbons, unruh} 
which is independent of the field equations of the theory. Later work also showed that the entropy attributed to a horizon, in sharp contrast, depends explicitly on the gravitational theory \cite{Wald, iyer&wald}.
Work in the last several decades attempted to understand the physical origin of the thermodynamical variables attributed to the horizons concentrating mostly on black hole horizons. In spite of extensive work and different possible suggestions for the source of, for example, entropy it is probably fair to say that we still do not quite understand the physics behind this phenomenon.

Somewhat ironically, the difficulty is not only about understanding gravitational entropy --- though that \textit{is} a serious issue ---  but also in relating it properly to the matter sources which can also possess an entropy described by conventional thermodynamics. In a situation involving both matter sources  and  gravity, it is not quite clear what precisely is the inter-relationship between the usual thermodynamic entropy of matter and the entropy of the horizon.  The two  extreme views which are possible would be: 
(i) The entropy of the horizon is the same as the entropy of the matter source, the gravity of which leads to the formation of the horizon (see, for e.g., \cite{israel, oppenheim}). 
(ii) The horizon entropy  arises from microscopic quantum structure of spacetime (see, for e.g., \cite{Padmanabhan:2009vy}) and the matter fields inherits the thermodynamic variables just as material kept in a hot oven inherits its temperature.  As usual, there are open questions in both points of view and it could very well be that neither approach fully captures the reality. There are at least two more facts which we need to keep in mind while studying a situation in which both normal  matter and gravity are present.

First, we  know that one of the thermodynamical variables, viz. the temperature, attributed to the vacuum state of the theory is observer dependent. In flat spacetime, an inertial observer would attribute zero temperature to the vacuum state while a uniformly accelerating Rindler observer will attribute a non-zero temperature to the vacuum state. This result continues to hold in an approximate sense \cite{dawood-detector, raval&hu} for more realistic non-inertial trajectories and indicates that vacuum fluctuations can mimic thermal fluctuations in certain contexts. Consider now a highly excited state of the vacuum with $n_{\bm k}$ quanta in a state labelled by the momentum ${\bm k}$. It is obvious that the thermodynamical properties of such a highly excited state [which could, for example, represent a quantum gas of particles] will also be viewed differently by inertial and non-inertial observers. In other words, once the thermodynamic properties of vacuum state have become observer dependent, the thermodynamic properties of \textit{any system} becomes observer dependent. We can no longer attribute to a box of gas a temperature or entropy, say, in an observer independent fashion. This situation is peculiar and requires deeper understanding.

Second, the co-existence of matter and gravitational field can lead to both kinematic and dynamical effects depending on whether we take into consideration the self-gravity of matter or not. For example, consider a spherical cloud of gas which collapses to form a black hole under its own gravity. It will be interesting to understand how the ``normal'' entropy of the gas cloud is related to the entropy of the black hole (see, for e.g., \cite{israel, oppenheim}), both with respect to an observer collapsing with the cloud and with respect to an outside observer (who will see the black hole formation only at asymptotic infinity). On the other hand, we can also ask a purely kinematical question of what happens to a cloud of gas, as regards its thermodynamical properties when it is located in a spacetime with horizon. In this case, the gas is \textit{not} self-gravitating and the horizon is produced by an external source or could even be due to the acceleration of the observer. 

This paper attempts to analyze the last aspect mentioned above, in the context of  the thermodynamical behaviour of a box of gas, treated as a  test system located  in an external spacetime with horizon, neglecting its self-gravity. Such a study is important to  distinguish sharply between the kinematical and dynamical effects when both matter and gravity are present. It will also reinforce the essential observer dependence of thermodynamics arising principly through Davies-Unruh effect.  
Specifically, we study in this paper a box containing a gas of indistinguishable particles in a  static background spacetime having a horizon. The box is oriented such that one of its faces is (approximately) parallel to the horizon surface (which we call the transverse direction) and its  proper `height' in the  direction normal to this face  is $K$ (borrowing the terminology from the case of approximate planar symmetry, we  call this the height of the box, which is in the direction of the gravitational acceleration). We then calculate the entropy of the gas when the box is situated far away and near  the horizon. We obtain the relevant expressions in two different ways. In Section \ref{stat mech}, we use the standard techniques of statistical mechanics  to find the thermodynamical quantities from the phase space volume of the system using the results of \cite{paddyphasespace}. We analyze its thermal behavior in various background spacetimes having a horizon for which the near horizon limit of the metric is the Rindler metric. Our analysis shows that far away from the horizon, the entropy of the gas depends on volume of the box with negligible corrections due to the background geometry. As we move towards the horizon, in general, entropy etc. depends on the location of  the box in a way dictated by the background metric (and in general is not a function of its volume alone because of finite size effects in an external geometry). In spherically symmetric spacetimes the transverse  directions  are unaffected and quantities such as transverse pressure take their usual Minkowski form. However, when the leading edge of the box is about \pl $L_p$  away from the horizon the entropy depends on a smaller volume $A_{\perp} L_p/2$ instead of the total volume $V$ of the box, upto an order $ {\cal O} (L_p/K)$. Further, we find that near the horizon all the thermodynamical quantities behave as though the box of gas has a volume $A_{\perp} L_p/2$ and is kept in an Minkowski spacetime. If we further assume that the box in thermal equilibrium is at the horizon temperature ${\beta_H}^{-1}$, then we get a $A_\perp / L_p^2$ dependance in entropy. In Section \ref{thermo}, we repeat our analysis using thermodynamics and obtain the same results as in section \ref{stat mech}. As a byproduct, we obtain the explicit expressions for the thermodynamic potentials in a general case of a spherically symmetric background metric of the form:
\begin{equation}
 ds^2 = -f(r)dt^2 + \frac{1}{f(r)} dr^2 + r^{2} d\Omega
\end{equation}
The conclusions are discussed in section \ref{conclu}.

\section{Thermodynamics of a box of gas}

We consider a box of an ideal gas consisting of $N$ indistinguishable classical Boltzmann particles to be in thermal equilibrium. We will assume that  the system does not backreact on the given background metric and study the thermal behavior of the system using two different approaches. First, we calculate the number of possible microstates of the system and  employ the usual techniques of statistical mechanics for a canonical ensemble to find all the thermodynamic variables from the partition function for the system. Second we write down the thermodynamical equations for the gas treated as an ideal fluid located in an external spacetime, solving which we find all the relevant thermodynamic quantities. Both approaches, of course, lead to the same results.

\subsection{Statistical Mechanics of the gas} \label{stat mech}

Consider the box of ideal gas  in thermal equilibrium with its surroundings at temperature $\beta^{-1}$. It is well known that the partition function of a system in canonical ensemble is sufficient to find all the thermodynamical variables of the system. The partition function $Z(\beta)$ is related by a Laplace transform to the density of states $g(E)$ available to the particles which, in turn, can be obtained from the phase space volume $P(E)$ as
\begin{eqnarray}
g(E) &=& \frac{dP(E)}{dE} \\
Q(\beta) &=& \int e^{-\beta E} g(E) dE = \int e^{-\beta E} d(P(E))  
\end{eqnarray}
Therefore, we only need a generally covariant expression for  the phase space volume $P(E)$ to work in a general co-ordinate system and determine the thermodynamics. Such a definition \cite{paddyphasespace} of phase space volume $P(E)$ in any static spacetime (described briefly in appendix[\ref{phasespacedef}] for completeness) is given by

\begin{equation}
P(E)=\int d^3x d^3p \ \Theta(E-\xi^a p_a)
\label{gdef1}
\end{equation}
where $p^a$ is the four-momentum of the particle and $\xi^a \equiv (1,\textbf{0})$ is the killing vector of the static spacetime. This expression of $P(E)$ is generally covariant, it is observer dependent due to the presence of the Killing vector $\xi^a$ used in defining the energy of the particles. When the spacetime has, say, two timelike Killing vectors in a region (for e.g. the ones corresponding to translation in Minkowski and Rindler time coordinates) used by two different class of observers, the phase space volume and hence the thermodynamic properties of the gas as measured by the two observers will be different. In $D+1$ dimensional spacetime, it is easy to show that the above relation reduces to 
\begin{equation}
{P}(E)=\frac{\pi^{D/2}}{\Gamma(\frac{D}{2}+1)} \int \sqrt{\gamma}d^Dx \left( \frac{E^2}{g_{00}}-m^2 \right)^{D/2} = \frac{\pi^{D/2}}{\Gamma(\frac{D}{2}+1)} \int d^Dx \sqrt{\gamma} \frac{E^D}{(g_{00})^{\frac{D}{2}}} \left(1 -\frac{m^2g_{00}}{E^2} \right)^{D/2}
\label{Dphdef}
\end{equation}
where $\gamma^{\alpha \beta}=g^{\alpha \beta}$ is the spatial part of the metric. From the second equality in the above equation, one can see that near a horizon where $g_{00} \rightarrow 0$, the contribution from the mass term can be ignored. Since we will be primarily interested in the near horizon behavior, we will take $m=0$ for the sake of calculational simplicity when dealing with curved spacetimes. We will later see that this assumption does not affect our key results as, near a horizon, the correction to the leading order term due to the non-zero mass turns out to be negligible. We will now explore the system in various backgrounds having a horizon. For having a standard reference to compare with, we begin with the Minkowski spacetime.

\subsubsection{Minkowski spacetime}

For $g_{ab} = \eta_{ab}$ in Eq.[\ref{Dphdef}], we find the usual form of the phase space volume $P(E)$
\begin{eqnarray}
P(E) &=& \frac{\pi^{D/2}}{\Gamma(\frac{D}{2}+1)} V E^D 
\longrightarrow \frac{4 \pi}{3} V E^3
\label{mphdef}
\end{eqnarray}
where we have taken $m=0$ and in the second step, we have taken $D=3$. The partition function $Q_1(\beta)$ for a single massless particle is then obtained to be
\begin{eqnarray}
Q_1(\beta) = \left[ \frac{D! \pi^{D/2}}{\Gamma(\frac{D}{2}+1)} \right] \frac{V_{D}}{ \beta^D} 
\longrightarrow 8 \pi \frac{V_{3}}{ \beta^3}
\label{3flatQ1}
\end{eqnarray}
where $V_D$ is the volume of the $D$ dimensional box. The partition function $Q_N(\beta)$ for $N$ non-interacting indistinguishable particles is $Q_N(\beta)=(1/N!)[Q_1(\beta)]^N$, where $(1/N!)$ is the Gibbs counting factor. For large $N$, we use the Stirling approximation $\log N! \sim N \log N -N$ to get the usual result:
\begin{equation}
\log{Q_N}=N\log{\left[ \left( \frac{D! \pi^{D/2}}{\Gamma(\frac{D}{2}+1)N} \right) \frac{V_{D}}{ \beta^D} \right]} -N
\end{equation}
The entropy of the gas is:
\begin{eqnarray}
S = \left(1- \beta \frac{\partial}{\partial \beta} \right)\log{Q_N} = N \left[ \log{\left(  \frac{D! \pi^{D/2}}{\Gamma(\frac{D}{2}+1)N}  \frac{V_{D}}{ \beta^D} \right)} + D\right] \longrightarrow N \left[ \log{\left(  \frac{8 \pi}{N}  \frac{V_{3}}{ \beta^3} \right)} + 3\right]
\label{mentropy}
\end{eqnarray}
which is the standard result \cite{pathria}. 
The corresponding equation of state obtained is the ideal gas equation:
\begin{eqnarray}
P = \frac{1}{\beta}\frac{\partial S}{\partial V} = \frac{N}{\beta V}
\label{minkowskipressure}
\end{eqnarray}
The total energy of the system is
\begin{equation}
E = -\frac{\partial\log{Q_N} }{\partial \beta} = \frac{DN}{\beta} \longrightarrow \frac{3N}{\beta}
\label{minkowskienergy}
\end{equation}
Since setting $m=0$ is effectively considering a relativistic gas, we get the correct factor of three in the above expression. From the Eq.[\ref{mentropy}] for entropy, one can see that entropy per particle $S/N$ depends on the logarithm of the total volume $V$ of the system. This is a standard result \cite{pathria} known in thermodynamics and is the expected of the behavior of entropy. It is this volume dependance of entropy that we investigate further when the system is in a background static spacetime having a horizon. We begin to do so starting with the simplest of static spacetimes with a horizon namely the Rindler spacetime.

\subsubsection{Rindler spacetime}
Consider the following form of the Rindler metric
\begin{eqnarray}
ds^2 &=&(1+\kappa x)^2dt^2-dx^2-dy^2-dz^2 \nonumber \\
&=&g_{00}(x_\parallel)dt^2-dx_\parallel^2-dx_\perp^2
\end{eqnarray}
given by the mapping 
\begin{equation}
X+\frac{1}{\kappa}=\kappa^{-1}(1+\kappa x)\cosh{\kappa t} \ ;\ \ T=\kappa^{-1}(1+\kappa x)\sinh{\kappa t}
\label{trans}
\end{equation}
from the Minkowski co-ordinates $(T,X)$ to the Rindler co-ordinates $(t,x)$. Here $x_\parallel \equiv x$ and $x_\perp \equiv (y,z)$. The Rindler observer's motion corresponds to a hyperbolic trajectory $(X+\frac{1}{\kappa})^2-T^2=\kappa^{-2}$  with constant proper acceleration $\kappa$ in the Minkowski frame. The Rindler observer perceives a horizon at $x=-\kappa^{-1}$  which is the $X-T-\kappa^{-1}=0$ surface in the Minkowski frame. Here $g_{00}=(1+\kappa x)^2$ and $\sqrt{\gamma}=1$. Hence the phase space volume for a single particle using Eqn.[\ref{Dphdef}] is
\begin{eqnarray}
P(E) &=& \frac{4}{3}\pi A_\perp \int_{x_a}^{x_b} dx \left[\frac{E^2}{(1+\kappa x)^2}-m^2\right]^{3/2} \nonumber \\
&=& \frac{4}{3}\pi E^3 A_\perp \int_{x_a}^{x_b} dx (1+\kappa x)^{-3}
\end{eqnarray}
where $A_\perp$ is the transverse area of the box and $(x_a,x_b)$ are the ends of the system in the longitudinal direction and we have set $m=0$ in the second step. On integrating, we get
\begin{eqnarray}
P(E) = \frac{2}{3}\pi E^3 A_\perp \frac{1}{\kappa}\left[ \frac{1}{(1+\kappa x_a)^2}-\frac{1}{(1+\kappa x_b)^2}\right]
= \frac{2}{3}\pi E^3 A_\perp \frac{1}{\kappa}\left[ \frac{1}{g_{00}(x_a)}-\frac{1}{g_{00}(x_b)}\right]
\label{grind}
\end{eqnarray}
One can check that in the $\kappa \rightarrow0$ limit, we get back the classical result Eqn.[\ref{mphdef}].

We now consider what happens when one edge of the box gets close to the horizon. Later on, we will treat the Rindler metric as an approximation close to the horizon of, for example, the Schwarzschild metric. In such a context, for an observer at infinity, the box of gas will take an infinite time to reach the horizon and will never cross the horizon. However,  given the inevitable quantum  uncertainty in the  position of any object \cite{planklength1}, and the fuzziness of the horizon, one cannot differentiate between a particle which is at a Planck length away from the horizon from the one which has crossed the horizon. Since the box of gas reaches $x_a =-\kappa^{-1}+L_p$ in large but yet a finite time, this is a more appropriate limit to consider for our purpose.
So consider what happens when one end of the box is at a Planck distance $L_p$ from the horizon $x_a =-\kappa^{-1}+L_p$. Let the height of the box be $H = x_b - x_a$, then $g_{00}(x_a) = \kappa^2L_p^2$ and  $g_{00}(x_b) = \kappa^2(H+L_p)^2$. (Here, and in what follows, one would prefer to work with proper distances, but since the spatial part of the metric is flat, that is $g_{\alpha \beta} = \delta_{\alpha \beta}$, the coordinate distances and proper distances are the same.) We then have for the phase space volume:
\begin{eqnarray}
P(E) &=& \frac{2}{3}\pi E^3 A_\perp \frac{1}{\kappa^3 L_p^2} \left(1 - \frac{L_p^2}{(H+L_p)^2} \right)
\label{ghor}
\end{eqnarray}
The one particle partition function is 
\begin{eqnarray}
 Q_1 =  \frac{4 \pi A_\perp}{\beta^3 \kappa^3 L_p^2} \left(1 - \frac{L_p^2}{(H+L_p)^2} \right) 
= 8 \pi \frac{ A_\perp (L_p/2)}{(\beta_{loc}(x_a))^3 } \left(1 - \frac{L_p^2}{(H+L_p)^2} \right)
\label{rindpart}
\end{eqnarray}
where $\beta_{loc}(x_a) = \beta \sqrt{g_{00}(x_a)} = \beta \kappa L_p$ is the redshifted local (Tolman) temperature \cite{tolman}. When $L_p/H \ll 1$ which is trivially satisfied for a finite height of the box, we can neglect the second term in the bracket which is of ${\cal O}(L_p^2 / H^2)$ compared to unity and obtain:
\begin{eqnarray}
 Q_1 \approx 8 \pi \frac{ A_\perp (L_p/2)}{(\beta_{loc}(x_a))^3 }
\label{rinQ1}
\end{eqnarray}
We can now calculate the various thermodynamic quantities from the $N$  particles partition function $Q_N=Q_1^N/N!$ in a  manner similar to what we did in the case of Minkowski spacetime. The entropy of the system is:
\begin{equation}
S \approx N \left[ \log{ \left(  \frac{8 \pi}{N}  \frac{A_\perp (L_p/2)}{ (\beta_{loc}(x_a))^3} \right)} + 3\right]
\label{rindlerentropy}
\end{equation}
(This expression is equivalent with the one  obtained in a completely different context in \cite{martinez})
The average energy of the system as measured by a stationary observer at $x_a$ is 
\begin{eqnarray}
U = -\frac{\partial \log Q_N}{\partial \beta_{loc}} 
= \frac{3N}{\beta_{loc}} 
\label{rindlerenergy}
\end{eqnarray}
To find the pressure, we can imagine the sides of box to be attached to movable pistons such that we can, one by one, displace each side by an infinitesimal amount while keeping the others fixed. The pressure is then related to the work done by the system against its change in volume and is given as $P= \beta_{loc}^{-1}\partial S/\partial V $. The transverse pressure is obtained when we do work on the box in one of the transverse directions keeping $x_{\parallel}$ and the remaining transverse coordinate fixed. We find that it obeys the usual ideal gas equation of state
\begin{eqnarray}
P_{\perp} &=& \frac{N}{\beta_{loc} V} 
\end{eqnarray}
showing that the transverse directions are not affected by the existence of the horizon.
Proceeding in a similar manner to find the pressure in the longitudinal direction at $x_a$, we again get the ideal gas equation of state but the pressure now depending on a much smaller volume $A_{\perp} (L_p/2)$ rather than the total volume $A_{\perp} K$ of the box
\begin{eqnarray}
P_{||x_a} = \frac{1}{\beta_{loc}}\frac{\partial(S)}{A_{\perp}\partial x_a} 
\approx \frac{N}{\beta_{loc} A_{\perp} (L_p/2)} 
\label{longpressurerindler}
\end{eqnarray}
Comparing the expressions of partition function, energy, longitudinal pressure and entropy in Eq.[\ref{rinQ1},\ref{rindlerenergy}, \ref{longpressurerindler},\ref{rindlerentropy}] with those in Minkowski spacetime in Eq.[\ref{3flatQ1},\ref{minkowskienergy},\ref{minkowskipressure},\ref{mentropy}] we find that near the horizon, the dynamics of the system is same as though the system was kept in a Minkowski spacetime but with the total volume $A_{\perp}H$ of the box replaced by a smaller volume $A_{\perp} (L_p/2)$  and with a inverse temperature $\beta = \beta_{loc}(x_a)$. 

The main reason for this effect is the large contribution to the phase space volume coming from near the horizon because of the infinite blueshift in the particle energy arising from the divergent nature of $g_{00}$ near the horizon. In other words, the density of states in phase space available to the particle near the horizon is huge as compared to the contribution from farther away. Thus the contribution to the entropy comes from only a fraction $ {\cal O} (L_p/H)$ of the microscopic matter degrees of freedom near to the horizon out of the total degrees present. Since $R_{abcd} =0$ for the Rindler spacetime, we see that even in a flat spacetime but having a horizon, the dependance of entropy is only on a small volume $A_{\perp} (L_p/2)$ instead of the whole volume $A_{\perp} H$ of the box. 

We therefore conclude that (i) this effect is purely kinematical in its origin and is independent of spacetime curvature (ii) it is observer dependent as the Rindler horizon is observer dependent. We will briefly comment on the second point before proceeding further. For an inertial observer (say the stationary observer at the origin in the Minkowski frame), the box of gas will appear to be accelerating. If we make an assumption, that at each point in its acceleration the gas maintains its thermal equilibrium, then Eq.[\ref{mentropy}] shows that the entropy as seen by the inertial observer depends on the volume of the box even as the box approaches the $X-T-\kappa^{-1}=0$ null surface. Here, the volume $V_M$ of the box as measured by the inertial observer is a function of time and is given by
\begin{equation}
 V_M = A_{\perp (M)} \left[ X_2(T) - X_1(T) \right]
\end{equation}
where $A_{\perp (M)}$ is the area of the box transverse to the $X$ direction and $X_2(T)$, $X_1(T)$ are the two ends of the box in the $X$ direction satisfying the hyperbolic trajectory equation 
\begin{equation}
X_1^2 - T^2 = \frac{(1 + \kappa x_a)^2}{\kappa^2} = L_p^2; \qquad X_2^2 - T^2 = \frac{(1 + \kappa x_b)^2}{\kappa^2} = (H + L_p)^2
\end{equation}
One can check that $X_2(T) - X_1(T)$ goes from $X_2(0) - X_1(0) =H $ at $T=0$ to $X_2 - X_1 \rightarrow 0 $ at $T \rightarrow \infty$ and hence the volume of the box goes from $V_M = A_{\perp (M)} H$ at $T=0$ to $V_M \rightarrow 0$ at $T \rightarrow \infty$. This is purely due to the continuous length contraction in the $X$ direction and is expected since the accelerated trajectory of the box is a result of continuous Lorentz boots in the $X-T$ plane. One should however note that this volume contraction in the dependence of entropy of the gas as it approaches the $X-T-\kappa^{-1}=0$ as $T \rightarrow \infty$ is completely different from the effect described above in which an area dependence is obtained.

It is also easy to check that if we include a mass term in the expression for phase space volume, the corrections to the leading term in the above expressions for partition function, entropy, etc are negligible and are of order ${\cal O} (L_pm)^2$.
Hence our assumption of taking $m=0$ does not affect our results in the order with which we are working and hold true to the same order for massive  particles as well. Further, since the box is fixed at its location near the horizon for arbitrarily large amount of time, it is reasonable to assume that the gas is thermalized at the same temperature as the horizon temperature. If we substitute for $\beta$ in the expression for entropy as $\beta = 2\pi /\kappa$ , then we get 
\begin{equation}
S \approx N \left[ \log{ \left(  \frac{4 \pi}{(2\pi)^3N}  \frac{A_\perp }{L_p^2} \right)} + 3\right]
\end{equation}
which is the familiar scaling of $A_\perp / L_p^2$. Of course, since by $L_p$ we strictly mean a length of the order of Planck length, the numerical factor of the order of unity cannot be fixed by this analysis.

\subsubsection{Schwarzschild Spacetime}

We next consider the system having its boundary in the shape of an 3-dimensional annular ring subtending a solid angle $\Omega$ at the origin and with an inner radius $r_a$ and outer radius $r_b$, located in 
the Schwarzschild spacetime with metric given by:
\begin{equation}
 ds^2 = -\left( 1-\frac{2M}{r}\right)dt^2 + \left(1- \frac{2M}{r} \right)^{-1} dr^2 + r^2 d\Omega^2
\end{equation}
The phase space volume Eq.[\ref{Dphdef}] becomes
\begin{eqnarray}
P(E)&=&\frac{4\pi E^3}{3} \int_{\Omega} \sin{\theta} \, d\theta d\phi \int_{r_a}^{r_b} \frac{r^2 dr}{\left(1- \frac{2M}{r} \right)^{2}} \nonumber \\
&=&\frac{4\pi \Omega E^3}{3} \biggl[ -\frac{(2M)^4}{r-2M} + 4(2M)^3 \log{(r-2M)} +6(2M)^2(r-2M) + 2(2M)(r-2M)^2 + \frac{(r-2M)^3}{3} \biggr]^{r_b}_{r_a} \nonumber \\
\label{schphspcexact}
\end{eqnarray}
We take the outer radius $r_b = 2M + H$ to be  such that $L_p \ll H \ll (2M)^2/L_p$; as regards the  inner radius $r_a = 2M + h$ which is  near the horizon, we will choose $h$ such that  the proper length from the horizon is equal to $L_p$. (This is precisely what we did in the previous section except that the proper and coordinate distances coincided in the Rindler spacetime.). This fixes $h$ to be:
\begin{eqnarray}
L_p = \int_{2M}^{h}\frac{dr}{\sqrt{1- \frac{2M}{r}} } 
\approx  2\sqrt{2hM}
\end{eqnarray}
where we have used $g_{00}(r) \approx (r-2M)g_{00}^{\prime}(2M) $ near the horizon. 
Then we have
\begin{eqnarray}
P(E) &=& \frac{4\pi E^3}{3} \Omega \frac{(2M)^4}{h} \biggl[1 - \frac{2h}{M} \log{\frac{h}{H}} + .. \biggr] \\
&\approx& \frac{4\pi E^3}{3} \Omega \frac{(2M)^4}{(L_p^2/8M)} \biggl[1 - \frac{L_p^2}{(2M)^2} \left( \log{\frac{h}{H}} - \frac{1}{12}\frac{H^3}{(2M)^3} + {\cal O}(H/2M)^2 \right) \nonumber  + {\cal O}(L_p/2M)^4 + .. \biggr] \nonumber \\
\end{eqnarray}
Ignoring the ${\cal O}(L_p/2M)^2$ and higher order terms; using $\kappa = g_{00}^\prime/2 = 1/4M$, we get
\begin{eqnarray}
P(E) \approx \frac{2\pi E^3}{3}  \frac{A_H}{(L_p^2 \kappa^3)}
\label{schphspc}
\end{eqnarray}
where $A_H = \Omega (2M)^2$ is the area of the horizon intercepted by solid angle $\Omega$.
To be precise, this is slightly different from the transverse area (facing the horizon) of $3$-dimensional annular container which is  $A_\perp = \Omega (2M + (L_p^2/8M))^2$. But they are equal to zeroth order of $(L_p/M)^2$ which will be the relevant order for us; so we will not make any distinction between the two. Also note that for a sufficiently large $H$, if we set the outer radius at $r_b = 2M + H$ and  let $H \longrightarrow \infty$, say, then the last term in Eq.[\ref{schphspcexact}] will also contribute to $P(E)$ at outer radius alongwith the term in Eq.[\ref{schphspc}] at inner radius. Since we are mainly interested in the effects due to a horizon and hence we do not consider such cases and treat $H$ to be bounded.
Calculating the partition function using Eq.[\ref{schphspc}], we get
\begin{eqnarray}
 Q_1 \approx  \frac{4 \pi A_H}{\beta^3 \kappa^3 L_p^2}  
\approx 4 \pi \frac{ A_H L_p}{\beta^3 (\sqrt{g_{00}(r_a)})^3}  
= 8 \pi \frac{ A_H (L_p/2)}{(\beta_{loc}(r_a))^3 }
\end{eqnarray}
The entropy of the system is then
\begin{equation}
S \approx N \left[ \log{ \left(  \frac{8 \pi}{N}  \frac{A_H (L_p/2)}{ (\beta_{loc}(r_a))^3} \right)} + 3\right]
\label{sentropy}
\end{equation}
The average energy of the system is 
\begin{eqnarray}
U = \frac{3N}{\beta_{loc}} 
\end{eqnarray}
The longitudinal pressure is 
\begin{eqnarray}
P_{r_a} &\approx& \frac{N}{\beta_{loc} A_{\perp} (L_p/2)} 
\end{eqnarray}
Once again we find that the near horizon dynamics of the system is reduced to the dynamics of a system in Minkowski spacetime with volume $V_3 =  A_H (L_p/2)$ and with $\beta = \beta_{loc}(r_a)$. This is in some sense anticipated since we know that the near horizon limit of the Schwarzschild metric is the Rindler metric and in the latter case we found the same thermodynamical behavior of the system. Also we know that far from the horizon at $R\rightarrow \infty$, the Schwarzschild metric reduces to the flat spacetime metric and the entropy of the gas depends on the total volume of the gas. Hence if we bring the box of gas from infinity to a distance $L_p$ from the horizon in a quasi-static way such that at each point it is in thermal equilibrium with its surroundings, then the dependance of entropy changes to $A_H (L_p/2)$. This behavior is again purely kinematical in its origin and is independent of curvature effects.
When the gas has come to a thermal equilibrium at the Hawking temperature of the horizon, we have $\beta = 2\pi /\kappa$ and the entropy becomes
\begin{equation}
S \approx N \left[ \log{ \left(  \frac{4 \pi}{(2\pi)^3N}  \frac{A_H }{L_p^2} \right)} + 3\right]
\end{equation}
showing again the $A_H/L_p^2$ scaling. In this case, it is of some interest to check the nature of the
 next higher order correction to the entropy. A simple calculation gives:
\begin{equation}
S \approx N \log{ \left[  \frac{4 \pi}{(2\pi)^3N} \left( \frac{A_H}{L_p^2} + \Omega \log{\frac{8HM}{L_p^2}} \right) \right]} + 3N
\end{equation}
Thus we get a correction to the area dependance of entropy by a term which has a $\log{M}$ dependence (rather than, say, a power law). It is curious to note that similar corrections of order $\log{M}$ to the black hole entropy of $A/4L_p^2$ have been proposed in the context of loop quantum gravity.

We now study a general metric with a horizon such that its near horizon limit is the Rindler metric. We show that the thermal behavior of the system near the horizon is similar to that in the Rindler spacetime.

\subsubsection{Spherically symmetric static spacetimes}

Consider the following form of the metric
\begin{equation}
 ds^2 = -f(r)dt^2 + \frac{1}{f(r)} dr^2 + r^{D-1} d\Omega^{D-1}
\label{dmetric}
\end{equation}
with the following conditions: $f(r_{{\cal H}}) = 0$ defines the horizon at $r=r_{{\cal H}}$ and $f^\prime(r_{{\cal H}}) = 2\kappa$ is non-zero and finite. The phase space volume is 
\begin{eqnarray}
P(E) &=& \frac{\pi^{D/2}}{\Gamma(\frac{D}{2}+1)} \int \frac{r^{D-1}}{\sqrt{f}} \, dr \ d\Omega^{D-1} \left(\frac{E^2}{f} - m^2\right)^{D/2} \\
&=& \frac{\pi^{D/2}}{\Gamma(\frac{D}{2}+1)} E^D \Omega \int^{r_b}_{r_a} \frac{r^{D-1}dr}{f^{\frac{D+1}{2}}}
\end{eqnarray}
where $\Omega$ is the corresponding solid angle in $D$ dimensions and we have taken $m=0$ in the last expression. Like in the previous cases we consider $r_a = r_{{\cal H}} + h$ to be very close to the horizon such that $h/r_{{\cal H}} \ll 1$ and $r_b = r_{{\cal H}} + H$ where $h$ and $H$ are co-ordinate distances of the ends of the ``box`` from the horizon along the radial direction. The main contribution to this integral comes from near the horizon, hence we evaluate the integral taking $g_{00} = f \approx (r-r_{{\cal H}})f^\prime(r_{{\cal H}})$. Also we use the proper length $L_p$ from the horizon instead of $h$ by using 
\begin{equation}
L_p = \int_{r_{{\cal H}}}^{h} dr/\sqrt{f} \approx 2\sqrt{h/f^\prime(r_{{\cal H}})}
\end{equation}   
The phase space volume then becomes
\begin{eqnarray}
P(E) &\approx& \frac{\pi^{D/2}}{\Gamma(\frac{D}{2}+1)} E^D \Omega \frac{1}{(f^\prime(r_{{\cal H}}))^{\frac{D+1}{2}}}\int \frac{r^{D-1}dr}{(r-r_{{\cal H}})^{\frac{D+1}{2}}}
\label{phasespaceinD}
\end{eqnarray}
Substituting $u = r-r_{{\cal H}}$ and expanding the numerator in a Binomial expansion, the integral becomes
\begin{eqnarray}
\int \frac{r^{D-1}}{(r-r_{{\cal H}})^{\frac{D+1}{2}}}dr=\int \frac{\left[u^{D-1} + ...+ (D-1)ur_{{\cal H}}^{D-2} + r_{{\cal H}}^{D-1} \right]}{u^{\frac{D+1}{2}}}du
\label{expan}
\end{eqnarray}
The main contribution to this integral comes from the lower limit of the integral of $r_{{\cal H}}^{D-1}/u^{\frac{D+1}{2}}$ near the horizon where $u = h$ and is obtained to be 
\begin{eqnarray}
\int \frac{ r_{{\cal H}}^{D-1} }{u^{\frac{D+1}{2}}}du \approx \frac{2r_{{\cal H}}^{D-1}}{(D-1)h^{\frac{D-1}{2}}}
\approx  \frac{2^D r_{{\cal H}}^{D-1}}{(D-1)L_p^{D-1}(f^\prime(r_{{\cal H}}))^{\frac{D-1}{2}}}
\end{eqnarray}
All the remaining terms contribute to the integral in Eq.[\ref{expan}] atmost by ${\cal O}(L_p/H)^2$ times the last term. Hence we can write to the lowest order:
\begin{eqnarray}
P(E) \approx \frac{\pi^{D/2}}{\Gamma(\frac{D}{2}+1)} E^D \Omega \frac{2^D r_{{\cal H}}^{D-1}}{(D-1)L_p^{D-1}(f^\prime(r_{{\cal H}}))^{D}} 
= \frac{\pi^{D/2}}{\Gamma(\frac{D}{2}+1)} E^D \frac{A_{D-1}}{(D-1)L_p^{D-1}\kappa^{D}}
\label{Dsphphspc}
\end{eqnarray}
where $A_{D-1} = \Omega r_{{\cal H}}^{D-1}$. The partition function becomes
\begin{eqnarray}
Q_1  \approx  \frac{1}{(D-1)}\frac{D! \pi^{D/2}}{\Gamma(\frac{D}{2}+1)} \left( \frac{A_{D-1}}{\beta^DL_p^{D-1} \kappa^{D}} \right) 
= \frac{D! \pi^{D/2}}{\Gamma(\frac{D}{2}+1)} \left( \frac{A_{D-1}(L_p/(D-1))}{(\beta_{loc}(r_a))^D} \right)
\end{eqnarray}
Comparing the above expression with the expression of partition function Eq.[\ref{rinQ1}] of the gas in Rindler spacetime, it is obvious that the thermal behavior of the gas in the two cases is identical. We calculate the entropy to be
\begin{eqnarray}
S &\approx& N \left[ \log{\left( \frac{D! \pi^{D/2}}{\Gamma(\frac{D}{2}+1)N}  \frac{A_{D-1}(L_p/(D-1))}{(\beta_{loc}(r_a))^D} \right)} + D\right]
\label{fentropy}
\end{eqnarray}
Once again we find the similar dependance of entropy on the volume $A_{D-1}L_p/(D-1)$.
When the gas is thermalized at the horizon temperature $\beta = 2\pi / \kappa$, the above expression reduces to
\begin{eqnarray}
S &\approx& N \left[ \log{\left( \frac{1}{(2\pi)^D (D-1)} \frac{D! \pi^{D/2}}{\Gamma(\frac{D}{2}+1)N}  \frac{A_{D-1}}{L_p^{D-1}} \right)} + D\right]
\end{eqnarray}

One should note that even when the metric given in Eq.[\ref{dmetric}] is a solution of the field equations for gravity in a general Lanczos-Lovelock theory \cite{Lanczos:1938sf, Lovelock:1971yv}, the form of the expressions obtained above for a box of gas will remain valid. The dependence of entropy changes from a volume scaling to an area scaling as we approach the horizon. We will comment more on this in section \ref{conclu}.

\subsection{Thermodynamics of the gas} \label{thermo}

We repeat the analysis of the previous section by doing thermodynamics  in a covariant manner \cite{mtw}. Basically, we find the same results as in the previous section but the analysis serves as a consistency check.

\subsubsection{\textbf{Formalism}}

Consider a box of ideal gas in thermal equilibrium with its surroundings in a given background metric. For a small volume element of the fluid, we can define the following quantities in the proper reference frame of the observer moving with the fluid element, that is, in the rest frame of the fluid element.
(i) $c^2\rho$ is the total energy density (including rest mass energy, thermal energy, etc). We have reintroduced $c$, the speed of light in vacuum and is not set equal to unity in the following analysis.
(ii) $p$ is the pressure
(iii) $n$ is the number density
(iv) $T$ is the temperature
(v) $s$ is the entropy per particle such that `$ns$' is the entropy density.
We take the energy momentum tensor of the gas satisfying $\nabla_bT^{ab}=0$ to be that of an ideal fluid:
\begin{eqnarray}
T^{ab} &=& (\rho + \frac{p}{c^2})u^au^b + pg^{ab} 
\label{tab}
\end{eqnarray} 
The first law of thermodynamics can be written, in terms of these variables as:
\begin{eqnarray}
d\rho &=& \frac{(\rho + \frac{p}{c^2})}{n}dn + nT ds
\label{1law}
\end{eqnarray}
We assume that the  two  usual constitutive equations will hold for a local inertial observer. First is the ideal gas equation
\begin{eqnarray}
p=nT
\label{idealgas}
\end{eqnarray}
and the second is the expression for the entropy density
\begin{eqnarray}
ns = n\log{\left[ \frac{p^{C_v}}{n^{C_p}}\right]} + nB
\label{entropydensity}
\end{eqnarray}
where $B$ is a constant, $C_v$ is the specific heat of the gas at constant volume, $C_p$ is the specific heat of the gas at constant pressure and  $C_p-C_v= 1$. Note that the above two equations are local in nature and are the equations which will describe an ideal gas.

Thus we have 9 unknowns, namely $\rho,p,n,T,s,u^a$ and 9 equations to solve for them, namely 4 equations from $\nabla_bT^{ab}=0$, first law of thermodynamics, two constitutive equations, normalization condition of proper velocity $u^au_a=-1$ and the baryon conservation law
\begin{equation}
\nabla_a(n u^a) = 0
\label{baryonconsv}
\end{equation}
We will now solve the equations for the gas in thermal equilibrium.

\subsubsection{\textbf{The solution}}

We follow the usual procedure of projecting the equation $\nabla_bT^{ab}=0$ along $u^a$ and orthogonal to $u^a$ with the latter accomplished by using the projection tensor $P^{ab} = g^{ab} + u^au^b/c^2$. The projection $u_a\nabla_bT^{ab}=0$ along $u^a$
 with the definition of energy momentum tensor in Eq.[\ref{tab}], gives
\begin{eqnarray}
\frac{d\rho}{d\tau} &=& \frac{(\rho + \frac{p}{c^2})}{n}\frac{dn}{d\tau}
\end{eqnarray}
Comparing with the first law in Eq.[\ref{1law}] we get $ds/d\tau =0$. Further the projection
 $P_a^c\nabla_bT^{ab}=0$ orthogonal to $u^a$ leads to the Euler equation
\begin{eqnarray}
(c^2\rho + p)a_i=P_{ik}\nabla^k p
\label{euler}
\end{eqnarray}
where $a^i = u^k\nabla_k u^i$. We will now specialize to the case of the gas  located in a spherically symmetric spacetime with metric
\begin{equation}
 ds^2 = -f(r)c^2dt^2 + \frac{1}{f(r)} dr^2 + r^{D-1} d\Omega^{D-1}
\end{equation}
The stationary nature of the equilibrium requires that:
\begin{eqnarray}
u^a = (\frac{1}{\sqrt{f}},0,0,0)
\end{eqnarray}
With the choice of our metric and the stationarity of $u^a$, all of the thermodynamic potentials are independent of time and the baryon conservation equation is satisfied identically. Further, the three Euler equations reduces to just one non-trivial equation describing force balance in the radial direction:
\begin{eqnarray}
\frac{dp}{dr}=-\frac{(c^2\rho + p)}{2f}\frac{df}{dr}
\label{eulerf}
\end{eqnarray}
The stationarity condition also implies that the local temperature obeys the Tolman condition\footnote{As an aside, note that, this well-known result can be obtained from the following argument: We know that for photons in a static metric $\omega\sqrt{-g_{00}}$ is a constant. Now consider two regions in the spacetime, say $A$ and $B$ which can exchange energy by the emission and absorption of photons. For thermal equilibrium to be maintained, the Planckian distribution (which depends on $\omega/T$) should always  hold both at $A$ and $B$. Hence, the temperatures should be related by $T_B\sqrt{-g_{00}(B)} = T_A\sqrt{-g_{00}(A)}$ which implies $T\sqrt{-g_{00}}$ is a constant as required.}.
\begin{eqnarray}
T\sqrt{f}=T_c
\label{tolman}
\end{eqnarray}
where $T_c$ is a constant usually identified as the temperature in the asymptotic limit $f\rightarrow 1$. Thus we are left with 5 unknowns $\rho,p,n,T,s$ and 5 equations namely first law of thermodynamics, two constitutive equations, one Euler equation and the Tolman condition. As we shall show, these can be solved even without assuming a particular form of $f(r)$. Using the two constitute equations in the first law, we get
\begin{eqnarray}
d\rho = \frac{\left( \rho - C_vp \right) }{n}dn + C_v dp
\end{eqnarray}
One can check that the above differential relation is defined locally at any arbitrary point and satisfies the integrability condition $\partial (\partial \rho /\partial n)/\partial p = \partial (\partial \rho /\partial p)/\partial n $ with both $\partial (\partial \rho /\partial n)/\partial p = n^{-1} (\partial \rho/\partial p - C_v) = 0 $ and $\partial (\partial \rho /\partial p)/\partial n = \partial (C_v)/\partial p = 0$.
We integrate the differential relation to find $\rho(n,p)$ at each $r$ which is kept fixed. The solution is
\begin{eqnarray}
\rho = An + C_v p
\end{eqnarray}
where $A = A(r)$ is a constant (i.e independent of $n,p$ while integrating) but can depend on $r$. The $r$ dependence can be easily determined from the fact that  the energy density should be
\begin{eqnarray}
c^2\rho = mc^2 n \sqrt{f} + C_vp
\label{energydensity}
\end{eqnarray}
(Note that in the Newtonian limit $\phi \ll c^2$, where $f=(1+2\phi/c^2)$,  this energy density becomes $c^2\rho = nmc^2 + m\phi  + C_vnT_c$ which is the sum of rest mass energy, gravitational potential energy and the thermal energy. This requires
 $A(r) = m\sqrt{f}$.) Using Eqns.[\ref{eulerf},\ref{idealgas},\ref{tolman},\ref{energydensity}], we get
\begin{eqnarray}
p=p_0 \left( f^{\frac{-C_p}{2}} \right) e^{-\frac{mc^2}{2T_c}f}
\label{pressure}
\end{eqnarray}
\begin{eqnarray}
n=\frac{p_0}{T_c} \left( f^{\frac{-C_v}{2}} \right) e^{-\frac{mc^2}{2T_c}f}
\label{number}
\end{eqnarray}
where $p_0$ is some constant of integration which we will fix by using the condition 
\begin{equation}
 \int ndV = N
\end{equation}
The total energy is given by:
\begin{eqnarray}
E = \int (\sqrt{f}c^2\rho) \frac{r^2\sin{\theta}}{\sqrt{f}}dr d\theta d\phi
\end{eqnarray}
This expression has the following interpretation. Since $c^2\rho(r)$ is the energy density as defined in the local rest frame of the fluid element, it is the local energy density as measured by the observer  at $r=constant$. Hence the redshifted energy density as measured by an observer at infinity is $\sqrt{f(r)}c^2\rho(r)$. (Here we may assume that the $\delta V$ associated with the energy density is same for both observers.) Hence the total energy as measured by an observer at infinity is the integral of $\sqrt{f}c^2\rho$ over the volume of the box.

The expressions for pressure, entropy, energy, etc match with their standard expressions in the literature when the gas is located in a flat spacetime or in a weak gravitational field; for example, the earth's atmosphere (see appendix \ref{variouslimits} for a discussion of different limits). We now discuss the case when the box of gas is near to the horizon.

\subsubsection{\textbf{Near horizon strong field limit}}

We take the form of $f(r)$ such that $f(r_{{\cal H}}) = 0$ defines the horizon at $r=r_{{\cal H}}$ and $f^\prime(r_{{\cal H}}) = 2\kappa$ is non-zero and finite.
It is then clear from the expressions for pressure Eq.[\ref{pressure}], number density Eq.[\ref{number}] and entropy density Eq.[\ref{entropydensity}], that the main contribution to them and to the integrated quantities such as total energy $E$ and total entropy $S$ comes from near the horizon because of the inverse dependance on $f$. The effective contribution from the factor $\exp(-\frac{mc^2}{2T_c}f)$ then is just unity near the horizon. So  we again set $m=0$ from the beginning as the corrections from the non-zero mass terms to the leading order terms are negligible. The pressure and number density then become
\begin{eqnarray}
p=p_0 \left( f^{-\frac{C_p}{2}} \right)
\end{eqnarray}
\begin{eqnarray}
n=\frac{p_0}{T_c} \left( f^{-\frac{C_v}{2}} \right)
\end{eqnarray}
Thus we see that the number density is very large near the horizon and starts to fall as $1/(r-r_{{\cal H}})^{C_v/2}$ in the outward direction. This is expected since the (coordinate) gravitational acceleration $a = f^\prime /2\sqrt{f}$ on a stationary fluid element is unbounded at the horizon. This result is consistent with the results of the previous section \ref{stat mech}, where we saw that the density of states in the phase space available to the particles, is huge near the horizon. Using the above expressions for pressure and number density in Eq.[\ref{entropydensity}], we find a similar behavior for the entropy density because of its linear dependance on number density
\begin{eqnarray}
ns = n\log{\left[ \frac{T^{C_p}}{p_0}\right]} + nB
\end{eqnarray}
Thus the entropy per particle $s$ is a constant and depends on the scaling of the constant $p_0$. To find $p_0$, we use the condition
\begin{eqnarray}
N =  \int \frac{4\pi r^2}{\sqrt{f}}ndr
= \frac{p_0}{T_c} \int \frac{4\pi r^2}{f^2}   dr 
\end{eqnarray}
where we have chosen $C_v =3$ for a ultra relativistic gas which is consistent with $m=0$. This integral has been evaluated before in Eqn.[\ref{schphspcexact}] and Eqn.[\ref{phasespaceinD}].  Hence, we get
\begin{eqnarray}
p_0 \approx \frac{NT_c}{A_{\perp}}2L_p^2 \kappa^3
\end{eqnarray}
where $L_p$ is the proper distance from the horizon to the nearer end of the box. The entropy density then becomes 
\begin{eqnarray}
ns \approx nB+n\log{ \left[ \frac{A_{\perp}L_p/2}{N\beta_{loc}^3} \right ] }
\end{eqnarray}
where $\beta = 1/T_c$. 
Thus we find that entropy per particle $s$ depends on the logarithm of the volume $A_{\perp}L_p/2$ and is consistent with the results obtained in the previous section. The total entropy obtained by integrating the above expression over the volume of the box is
\begin{eqnarray}
S \approx NB+N\log{ \left[ \frac{A_{\perp}L_p/2}{N\beta_{loc}^3} \right ] }
\end{eqnarray}
Substituting $p_0$ in the expression of pressure and evaluating it at distance $L_p$, we get
\begin{eqnarray}
p_{L_p} &\approx& \frac{N}{\beta_{loc}A_{\perp}L_p/2}
\end{eqnarray}
The total energy is 
\begin{eqnarray}
E = \int (\sqrt{f}c^2\rho) \frac{r^2\sin{\theta}}{\sqrt{f}}dr d\theta d\phi = C_v NT_c
\end{eqnarray}
The above expressions of pressure, entropy and energy are essentially same as the corresponding expressions obtained in the previous section using the techniques of statistical mechanics.

A simple way to understand these results is to recall the expression of entropy of an isothermal ideal gas of N particles, each of mass $m$ located in a uniform weak gravitational field $g$, say on the surface of the earth. The expression obtained e.g. in appendix[\ref{earthsurface}], is
 \begin{eqnarray}
S = N(B+1) - \left( \frac{NmgL/T_c}{e^{mgL/T_c}-1} \right)  + N\log{\left[ \frac{(T_c)^{C_p}A_{\perp}}{Nmg}\left( \frac{e^{mgL/T_c}-1}{e^{mgL/T_c}} \right) \right]}
\end{eqnarray}
There are two length scales (i) $L$, the length of the box and (ii) $\lambda = T_c/mg$ in the above expression. The ratio $L/\lambda$ can be thought of as a measure of the relative strength between the gravitational potential energy and the thermal energy. When $\lambda \gg L$, the logarithmic dependence in the second term becomes $\log{\left(T_c^{C_V}A_{\perp}L \right)}$ which is the usual `volume of the box' dependence of entropy we are familiar with. However, if $L \gg \lambda$, the logarithmic dependence in the second term changes to $\log{\left( T_c^{C_V}A_{\perp}\lambda \right)}$ showing that the entropy now depends on a smaller volume than the total volume of the box and is analogous to the behavior of entropy of the gas near the horizon. This would correspond to a box with vertical length $L$ that is bigger than the scale length $\lambda \propto T/g$ of the atmosphere, which is unrealistic in normal circumstances. But note that if $g\rightarrow \infty$, then $\lambda \rightarrow 0$ for any finite $L$ and we get the area dependence. This is what happens in the case of a box approaching the horizon. When the box of gas is near to the horizon the two length scales involved are (i) $a \sim 1/L_p$, the gravitational acceleration near the horizon and (ii) $H$, the height of the box. Since $H$ is a macroscopic quantity, it can never be smaller than $L_p$ and hence we always have $H \gg a^{-1}$ near the horizon which gives rise to the $\log{\left( T_c^{C_V}A_{\perp}L_p \right)}$ dependence of entropy. When one doubles the system by making a scaling change $L_i \rightarrow \gamma L_i$ for $i=1,2.. ,D$ and $N \rightarrow \gamma^3 N$ where $\gamma^3 =2$, the entropy of the gas near the horizon goes as $S \rightarrow \gamma^3 N \log{A_{\perp}L_p/\gamma N}$ instead of the usual $S \rightarrow \gamma^3 N \log{V/ N} = \gamma^3 S $. Thus the extensivity property of entropy no longer holds and one can check that it does not hold even in the weak field limit discussed above when $L \gg \lambda$ that is, when gravitational effects subdue the thermal effects along the direction of the gravitational field.

\section{Discussion and Conclusions}\label{conclu}

To summarize, we have studied in two independent ways, the kinematic relationship between entropy and area near a Rindler-like horizon in a static spacetime. We have shown that when the box is far away from the horizon, the entropy of the gas depends on the volume of the box in the asymptotic flat limit of the background spacetime, except for small corrections due to background geometry. As the box is moved closer to the horizon with one (leading) edge of the box at about Planck length ($L_p$) away from the horizon, the entropy shows an area dependence rather than a volume dependence. More precisely, it depends on a small volume $A_{\perp} L_p/2$ of the box, upto an order $ {\cal O} (L_p/K)^2$ where $A_{\perp}$ is the transverse area of the box and $K$ is the longitudinal size of the  box related to the distance between leading and trailing edge in the vertical direction (i.e the of the gravitational field). Thus the contribution to the entropy comes from only a fraction ${\cal O} (L_p/K)$ of the matter degrees of freedom and the rest are suppressed when the box approaches the horizon. Near the horizon all the thermodynamical quantities behave as though the box of gas has a volume $A_{\perp} L_p/2$ and is kept in a Minkowski spacetime. Since all these effects are true in a Rindler spacetime, we argued that they are purely kinematic in their origin and independent of spacetime geometry. When the equilibrium temperature of the gas is taken to be equal to the the horizon temperature, we got the familiar $A_\perp / L_p^2$ dependance in the expression for entropy. All these results hold in a $D+1$ dimensional spherically symmetric spacetime.

We saw that the expression of entropy of a gas changes from a volume dependence to an area dependence as we change our perspective from that of an inertial observer to an accelerated observer highlighting the role of observer dependence in the thermal behavior of a gas. Such an observer dependence is well-known in the familiar \textit{quantum} phenomenon in which the inertial vacuum state of a scalar field in a flat spacetime appears to be a thermal state to a Rindler observer. If we treat normal matter systems as highly excited states of the vacuum we do expect an observer dependence for all thermodynamical variables. Here, the real surprise is probably in the fact that we have done a purely \textit{classical} analysis. The phenomenon seems to be closer to another well-known classical fact, viz. that an observer outside a horizon sees material systems to reach the horizon only after infinite time while the comoving observer will cross the horizon in finite time as shown by the comoving clock. This is not usually christened explicitly as `observer dependence' in literature but it definitely is. Most of the effects studied in this paper are related to the vanishing of the lapse function on the horizon and the consequent infinite redshift \cite{TPmag}. Ultimately, this is related to the existence of two different time coordinates adapted to two different observers. (In the context of inertial and Rindler frames, these are the $\partial/\partial T$ and $\partial/\partial t$.) Since the standard observer dependent temperature etc of the vacuum state is also related to this feature, viz. the non-linear relationship between the two time coordinates, one may claim that the root cause of both the classical observer dependence studied here and the quantum observer dependence of the temperature of the vacuum state are the same.

There is another aspect of our analysis which is worth noting. In maintaining the validity of the generalized second law, which was the original concern and motivation for Bekenstein's proposal, the loss of the entropy of an object falling into the horizon is related to the increase of the entropy of the horizon so that the total entropy never decreases. In the conventional analysis, usually done in the case of black holes, one considers the loss of the \textit{energy} of the object  and associates  with it the increase in the mass of the black hole and hence the area of the black hole. However as viewed from an outside observer  the infalling matter takes  infinite time to reach the horizon and hence it never crosses the horizon. One could therefore claim that no entropy is lost. This we believe is incorrect because one cannot operationally distinguish between a particle located within one Planck length from the horizon and a particle which has crossed the horizon and the former event occurs in large but yet a finite time. Given such an operational interpretation, it is interesting to study the matter entropy directly.  We see that although the degrees of freedom contributing to entropy are expected to be distributed all over the volume of the object, near the horizon only those degrees contribute which are in a small volume $A_{\perp} L_p/2$ and will interact with the horizon degrees of freedom. 

Finally, note that we had not assumed any gravitational field equations obeyed by the  spherically symmetric metric given in Eq.[\ref{dmetric}]. It could, for example, be a solution of the field equations for gravity in a very general theory like e.g., Lanczos-Lovelock theory \cite{Lanczos:1938sf, Lovelock:1971yv}. Our analysis of the thermal behaviour of box of gas will still remain valid and scaling change of the entropy volume dependence to an area dependence, as we approach the horizon, will hold. However, for a general theory like e.g., Lanczos-Lovelock theory in $D+1$ dimensions ($D > 4$), the entropy of a black hole  is not proportional to the $D-1$ dimensional hypersurface-area of the horizon. It follows that when both matter sources and gravity are present, the gravitational entropy is dependent on the theory of gravity whereas the matter entropy is independent of the field equations of gravity \cite{TPqglesson}. This is similar to the result in the case of entanglement entropy and it is not clear whether the   regularization procedure of introducing a cutoff of order Planck length near the horizon should be different, say, in the case of Lanczos-Lovelock theory (see e.g., \cite{TPentangle}). Alternatively, it could be that the entropy of matter degrees of freedom and that of gravitational degrees of freedom scale differently in general theories and the area scaling we obtain for the gravitational entropy is a result of no special significance.
These issue requires further study.

\section*{Acknowledgements}
The authors thank D.Kothawala for useful discussions and comments on the
draft. SK is supported by a Fellowship from the Council of Scientific and Industrial Research (CSIR), India.

\appendix{}

\section*{Appendix:}

\section{Phase space volume}\label{phasespacedef}

To work in a general co-ordinate system, we need to have an invariant description of the phase space volume $P(E)$. We follow the definition of phase space volume $P(E)$ as given in \cite{paddyphasespace} which we will describe  here in brief.

To define $P(E)$ one first needs to have a conserved definition of energy $E$. This is done with the help of the timelike killing vector of the static spacetime $\xi^a=(1,\textbf{0})$. A conserved energy $E$ is defined through $E=\xi^a p_a$ where $p^a$ is the four-momentum of the particle. The phase space volume is taken to be
\begin{equation}
P(E)=\int d^3x d^3p \ \Theta(E-\xi^a p_a)
\end{equation}
The product $d^3x d^3p$ can easily be shown to be invariant. Hence the above expression for $P(E)$ is invariant.
Using the fact $p^ap_a=m^2$, one writes $\xi^a p_a=g_{00}p^0=g_{00}^{1/2}(m^2+\gamma^{\alpha \beta}p_\alpha p_\beta)^{1/2}$, where $\gamma^{\alpha \beta}=g^{\alpha \beta}$ is the spatial part of the metric. The $p$ integration will give the volume of a 3-sphere in momentum space of radius $(E^2/g_{00}-m^2)^{1/2}$ which is $4 \pi/3 \sqrt{\gamma} (E^2/g_{00}-m^2)^{3/2}$
Therefore, the phase space volume takes the form
\begin{equation}
P(E)=\frac{4}{3}\pi \int \sqrt{\gamma}d^3x (E^2/g_{00}-m^2)^{3/2}
\label{phdef}
\end{equation}
In $D+1$ dimensional spacetime, it is easy to show that
\begin{equation}
{P}(E)=\frac{\pi^{D/2}}{\Gamma(\frac{D}{2}+1)} \int \sqrt{\gamma}d^Dx (E^2/g_{00}-m^2)^{D/2}
\end{equation}
One way to interpret Eq.[\ref{phdef}] as shown in \cite{paddyphasespace} is to define the energy measured by a locally defined observer $u^a$ as $E_{loc}=u^ap_a=(\xi^ap_a)g_{00}^{-1/2}=Eg_{00}^{-1/2}$. Then one can treat $\left((E/\sqrt{g_{00}})^2-m^2\right)^{1/2}$ as the local momentum $p_{loc}$. Then
\begin{equation}
P(E)=\frac{4}{3}\pi \int \sqrt{\gamma}d^3x \left[p_{loc}(x^\alpha)\right]^3
\end{equation}
Hence one gets the integral in $P(E)$ as the integral of local momentum over the whole volume of the system, which may not be proportional to the volume of the system. In Minkowski spacetime with the metric $\eta_{ab}$, the momentum $p_{loc}$ is a constant $p$ and one gets back the volume $V = \int \sqrt{\gamma}d^3x$ and the phase space volume takes the usual form for a single free particle moving at relativistic momentum $p=(E^2-m^2)^{1/2}$
\begin{equation}
P(E)=\frac{4}{3}\pi V [p]^3 = \frac{4}{3}\pi V [E^2-m^2]^{3/2}
\end{equation}

\section{Various limits}\label{variouslimits}
\subsection{Flat spacetime}
In this case $f=1$ and we get back our usual results in flat spacetime.
The ideal gas equation
\begin{eqnarray}
PV=NT
\end{eqnarray}
The total energy and entropy of the system is
\begin{eqnarray}
E = Nmc^2 + C_vNT
\end{eqnarray}
\begin{eqnarray}
S= NB + N\log{\frac{T^{C_p}}{p}}
\end{eqnarray}

\subsection{Weak field limit: $R_0 \gg h$} 
Consider the box of the gas to be on the surface of the earth (say). Let the radius of the surface of the earth be denoted by $R_0$ and the height measured from the surface be $h$ such that $R_0 \gg h$. Then we have
\begin{eqnarray}
f &=& \left( 1 - \frac{2GM}{c^2(R_0+h)} \right) \nonumber \\
&\approx& 1 - \frac{2GM}{c^2R_0} + \frac{2GMh}{c^2R_0^2}
\end{eqnarray}
Then the pressure can be approximated as 
\begin{eqnarray}
p &\approx& p_0 \left( f^{\frac{-C_p}{2}} \right)  e^{-\frac{mc^2}{2T_c} \left(1- \frac{2GM}{R_0}\right)} e^{-\frac{mgh}{T_c}} \\
&\approx& p_0 \left( 1 - \frac{2GM}{c^2R_0} + \frac{2gh}{c^2}\right)^{\frac{-C_p}{2}}   e^{-\frac{mc^2}{2T_c} \left(1- \frac{2GM}{R_0}\right)} e^{-\frac{mgh}{T_c}}
\end{eqnarray}
where $g= GM/R_0^2$ is the gravitational acceleration near the earth's surface. Similarly the number density is 
\begin{eqnarray}
n &\approx& \frac{p_0}{T_c} \left( f^{\frac{-C_v}{2}} \right)  e^{-\frac{mc^2}{2T_c} \left(1- \frac{2GM}{R_0}\right)} e^{-\frac{mgh}{T_c}} \\
&\approx& \frac{p_0}{T_c} \left( 1 - \frac{2GM}{c^2R_0} + \frac{2gh}{c^2}\right)^{\frac{-C_v}{2}}   e^{-\frac{mc^2}{2T_c} \left(1- \frac{2GM}{R_0}\right)} e^{-\frac{mgh}{T_c}}
\end{eqnarray}

\subsection{Non-relativistic weak field limit: $\phi \ll c^2$ , $R_0 \gg h$} \label{earthsurface}
If we further consider the gas on the surface of the earth to be non-relativistic and take the limit $(\phi = -2GM/R)\ll c^2$ in addition to $R_0 \gg h$, then the power in $f$ appearing in the expressions for pressure and number density in the above case can be neglected. Then we get
\begin{eqnarray}
p &\approx& \bar{p_0} e^{-\frac{mgh}{T_c}} 
\end{eqnarray}
and
\begin{eqnarray}
n &\approx& \frac{\bar{p_0}}{T_c} e^{-\frac{mgh}{T_c}} 
\end{eqnarray}
Therefore on normalizing, we get
\begin{eqnarray}
\bar{p_0} = \int n dV = \frac{Nmg}{A} \left( \frac{e^{mgL/T_c}}{e^{mgL/T_c}-1} \right)
\end{eqnarray}
The energy is 
\begin{eqnarray}
E = Nmc^2 + C_pNT - \left( \frac{NmgL}{e^{mgL/T_c}-1} \right) - \frac{mNGM}{R_0}
\end{eqnarray}
The total entropy is then
\begin{eqnarray}
S = N(B+1) - \left( \frac{NmgL/T_c}{e^{mgL/T_c}-1} \right)  + N\log{\left[ \frac{(T_c)^{C_p}A_{\perp}}{Nmg}\left( \frac{e^{mgL/T_c}-1}{e^{mgL/T_c}} \right) \right]}
\end{eqnarray}
For the sake of comparison, we give below the standard expressions of energy and entropy upto a constant \cite{sears}. One can see that both agree upto a constant.
\begin{eqnarray}
E = \frac{5}{2}NT - \left( \frac{NmgL}{e^{mgL/T}-1} \right)
\end{eqnarray}
\begin{eqnarray}
S = \frac{E}{T}  + N\log{\left[ \frac{(T)^{5/2}A_{\perp}}{Nmg}\left( \frac{e^{mgL/T}-1}{e^{mgL/T}} \right) \right]}
\end{eqnarray}


\begin{thebibliography}{25}
\providecommand{\natexlab}[1]{#1}
\providecommand{\url}[1]{\texttt{#1}}
\expandafter\ifx\csname urlstyle\endcsname\relax
  \providecommand{\doi}[1]{doi: #1}\else
  \providecommand{\doi}{doi: \begingroup \urlstyle{rm}\Url}\fi

\bibitem{bekentropy}
J.~D. Bekenstein.
\newblock \emph{Nuovo Cim. Lett.}, \textbf{4}, \penalty0 737-740, 1972.

\bibitem{bekentropy2}
Jacob~D. Bekenstein.
\newblock \emph{Phys. Rev.},\textbf{ D9}, \penalty0 3292-3300, 1974.

\bibitem{bekentropy3}
Jacob~D. Bekenstein.
\newblock \emph{Phys. Rev. D}, \textbf{7}, \penalty0 2333-2346, 1973.

\bibitem{hawking}
S.~W. Hawking.
\newblock \emph{Commun. Math. Phys.}, \textbf{43}, \penalty0 199-220, 1975.

\bibitem{hawking2}
S.~W. Hawking.
\newblock \emph{Phys. Rev. D}, \textbf{13}, \penalty0 191-197, 1976.

\bibitem{Davies}
P.~C.~W. Davies.
\newblock \emph{J. Phys.}, \textbf{A8}, \penalty0 609-616, 1975.

\bibitem{gibbons}
G.~W. Gibbons and S.~W. Hawking.
\newblock \emph{Phys. Rev. D}, \textbf{15}, \penalty0 2738-2751, 1977.

\bibitem{unruh}
W.~G. Unruh.
\newblock \emph{Phys. Rev. D}, \textbf{14}, \penalty0 870--892, 1976.

\bibitem{Wald}
Robert~M. Wald.
\newblock \emph{Phys. Rev.}, \textit{D48}, \penalty0 3427-3431, 1993.

\bibitem{iyer&wald}
Vivek Iyer and Robert~M. Wald.
\newblock \emph{Phys. Rev. D}, \textbf{50}, \penalty0 846-864, 1994.

\bibitem{israel}
F.~Pretorius, D.~Vollick, and W.~Israel.
\newblock \emph{Phys. Rev.}, \textbf{D57},\penalty0 6311-6316, 1998.

\bibitem{oppenheim}
Jonathan Oppenheim.
\newblock \emph{Phys. Rev.}, \textbf{D65},\penalty0 024020, 2002.

\bibitem{Padmanabhan:2009vy}
Padmanabhan T 2010 \textit{Mod. Phys. Lett.} \textbf{A 25} 1129  [arXiv:0912.3165];
 \textit{Phys. Rev.} \textbf{D 81} 124040 (2010) [arXiv:1003.5665]; for a review see,
T.~Padmanabhan.
\newblock \emph{Rept. Prog. Phys.}, \textbf{73},\penalty0 046901, 2010.

\bibitem{dawood-detector}
Dawood Kothawala and T.~Padmanabhan.
\newblock \emph{Phys. Lett.}, \textbf{B690},\penalty0 201-206, 2010.

\bibitem{raval&hu}
Alpan Raval, B.~L. Hu, and Don Koks.
\newblock \emph{Phys. Rev. D}, \textbf{55},\penalty0 4795-4812, 1997.

\bibitem{paddyphasespace}
T.~{Padmanabhan}.
\newblock \emph{Physics Letters A}, \textbf{136},\penalty0 203--205, 1989.

\bibitem{pathria}
R.~K. Pathria.
\newblock \emph{{Statistical Mechanics}}.
\newblock Elsevier, 2005.


\bibitem{planklength1}
T.~{Padmanabhan}, 1997 \textit{Phys. Rev. Letts} \textbf{78} 1854 [hep-th-9608182];
                   \textit{Phys. Rev.} \textbf{D57} (1998)  6206 ;
               \textit{Class. Quan. Grav.}, \textbf{4}, (1987) L107;
\newblock \emph{Annals of Physics}, 165:\penalty0 38--58, November
  1985{\natexlab{a}}.


\bibitem{tolman}
Richard~C. Tolman and Paul Ehrenfest.
\newblock \emph{Phys. Rev.}, \textbf{36},\penalty0 1791-1798, 1930.

\bibitem{martinez}
D.~J. {Louis-Martinez}.
\newblock {\textit{Classical relativistic ideal gas in thermodynamic equilibrium in a
  uniformly accelerated reference frame}}.
\newblock ArXiv:1012.3063, 2010.

\bibitem{Lanczos:1938sf}
Cornelius Lanczos.
\newblock \emph{Annals Math.}, \textbf{39}, \penalty0 842--850, 1938.

\bibitem{Lovelock:1971yv}
D.~Lovelock.
\newblock \emph{J. Math. Phys.}, \textbf{12}, \penalty0 498--501, 1971.
\newblock \doi{10.1063/1.1665613}.

\bibitem{mtw}
Misner, Thorne, Wheeler, \textit{Gravitation}, W. H. Freeman and Company (1973).


\bibitem{TPmag}
 T.Padmanabhan, \textit{Phys. Rev. Letts}.,  \textbf{81}, 4297 (1998) [hep-th-9801015];
 T.Padmanabhan, \textit{Phys. Rev.}, \textbf{D59}, 124012 (1999) [hep-th-9801138].

\bibitem{TPqglesson}
T. Padmanabhan,  \textit{Lessons from Classical Gravity about the Quantum Structure of Spacetime}, [arXiv:1012.4476]

 
\bibitem{TPentangle} 
Padmanabhan T  2010 \textit{Phys. Rev. D} \textbf{82} 124025
[arXiv:1007.5066].

\bibitem{sears}
Sears and Salinger, \textit{Thermodynamics, Kinetic Theory, And Statistical Thermodynamics}, Addison Wesley Longman (1975).

\end{thebibliography}
\end{document}